\documentclass{bmcart}

\usepackage[utf8]{inputenc} 
\usepackage{amsmath,mathtools}
\usepackage{algorithm}
\usepackage[noend]{algpseudocode}
\usepackage[colorinlistoftodos]{todonotes}
\usepackage{graphicx,graphics,subfigure,multicol} 
\usepackage{url}
\usepackage{booktabs}
\usepackage{multirow}
\usepackage{threeparttable}
\usepackage{tabularx} 
\usepackage{todonotes}
\usepackage{ulem}
\usepackage{lscape}
\usepackage{wasysym}

\startlocaldefs
\endlocaldefs

\begin{document}

\begin{frontmatter}

\begin{fmbox}
\dochead{Research}

\title{Coupling the circadian rhythms of population movement and the immune system in infectious disease modeling}

\author[addressref={aff1}]{\inits{ZD}\fnm{Zhanwei} \snm{Du}}
\author[addressref={aff2},email={holme@pi.cns.titech.ac.jp}]{\inits{PH}\fnm{Petter} \snm{Holme}}
\address[id=aff1]{\orgname{Department of Integrative Biology, The University of Texas at Austin}, \postcode{78712}, \city{Austin TX}, \cny{United States of America}}
\address[id=aff2]{\orgname{Tokyo Tech World Research Hub Initiative (WRHI), Institute of Innovative Research, Tokyo Institute of Technology}, \postcode{226-8503}, \city{Yokohama}, \cny{Japan}}


\end{fmbox}
\begin{abstractbox}

\begin{abstract}
\parttitle{Background}
The dynamics of infectious diseases propagating in populations depends both on human interaction patterns, the contagion process and the pathogenesis within hosts. The immune system follows a circadian rhythm and, consequently, the chance of getting infected varies with the time of day an individual is exposed to the pathogen. The movement and interaction of people also follow 24-hour cycles, which couples these two phenomena. 

\parttitle{Methods}
We use a stochastic metapopulation model informed by hourly mobility data for two medium-sized Chinese cities. By this setup, we investigate how the epidemic risk depends on the difference of the clocks governing the population movement and the immune systems. 

\parttitle{Results}
In most of the scenarios we test, we observe circadian rhythms would constrain the pace and extent of disease emergence. 
The three measures (strength, outward transmission risk and introduction risk) are highly correlated with each other. For example of the Yushu City, outward transmission risk and introduction risk are correlated with a Pearson's correlation coefficient of 0.83, and the risks correlate to strength with coefficients of $-0.85$ and $-0.75$, respectively (all have $p<0.05$), in simulations with no circadian effect and $R_0=1.5$.

\parttitle{Conclusions}
The relation between the circadian rhythms of the immune system and daily routines in human mobility can affect the pace and extent of infectious disease spreading. Shifting commuting times could mitigate the emergence of outbreaks.

\end{abstract}


\end{abstractbox}
%

\end{frontmatter}

\section*{Background}

Infectious diseases are a major burden to global health and will continue to be that for the foreseeable future. They hinder us in our everyday lives, and thus stops society from operating at well as it could. There is also a looming threat of pandemic of novel and hard-to-cure pathogens. There are both positive and negative factors affecting the long-term threats of infectious diseases---the continuous advances of medicine are a factor that keeps reducing the threat. Factors working in the opposite directions include: the increase of drug resistant pathogens, and urbanization. Since a few years back, most people in the world live in cities~\cite{cohen2003human}, and the fraction of city dwellers keep increasing. Bringing people together like a city does, increases the rate people meet close enough to transmit infectious diseases~\cite{merler2011determinants}. This makes epidemic thresholds lower, outbreaks more frequent and faster. For these reasons, stopping the epidemic spreading is almost synonymous with stopping them in cities.

Disease spreading is a phenomenon that is affected by a multitude of factors. Some of these occurs at different time scales---like evolution of pathogens is usually slower than the microscopic pathogenesis in individuals, that typically is slower than the mixing and interaction of people. In such a case, in epidemic modeling, one can separate the timescales and treat the slower changing feature as static. In other cases, the timescales are similar and then one cannot simplify a model like that. The life in cities follow a 24-hour, circadian rhythm. Many people commute at the same times every day~\cite{zhong2015measuring,amini2014impact,du2019inter}. People using public transport could be very close together at the rush hours of the day. The immune systems of people also follow a circadian rhythm. Quoting Ref.~\cite{Man999}: ``Nearly every arm of the immune response (innate and adaptive) has been reported to oscillate in a circadian manner.'' The best evidence comes from studies of herpes and influenza virus infection associated with the circadian clock~\cite{edgar2016cell}. Typically, the immune system is most effective in the midnight~\cite{Bass994,scheiermann2018clocking}. This means that the synchronization of the travel and the immune systems could affect the predictions of disease models and should ideally not be ignored. Still, to the best of our knowledge, no study takes this coupling into account but assume epidemiological characteristics (i.e., the transmission rate) are static over the course of hours and days~\cite{hoen2015epidemic,zhu2016inferring}.

If the immune system and the travel patterns were the only phenomena following circadian patterns, then taking their coupling into account would be straightforward. One could simulate the disease propagation at a time resolution of days using standard models, only the parameter values would need to be adjusted. However, this is not the case. Many types of infections have a faster development than could be captured by such a coarse time resolution~\cite{du2018periodicity}. Thus, for accurate modeling one need to simultaneously take the travel pattern, the strength of the immune systems, and the disease propagation into account. This is what we set out to do in this paper.

\begin{figure}[!ht]
\includegraphics[width=0.9\textwidth]{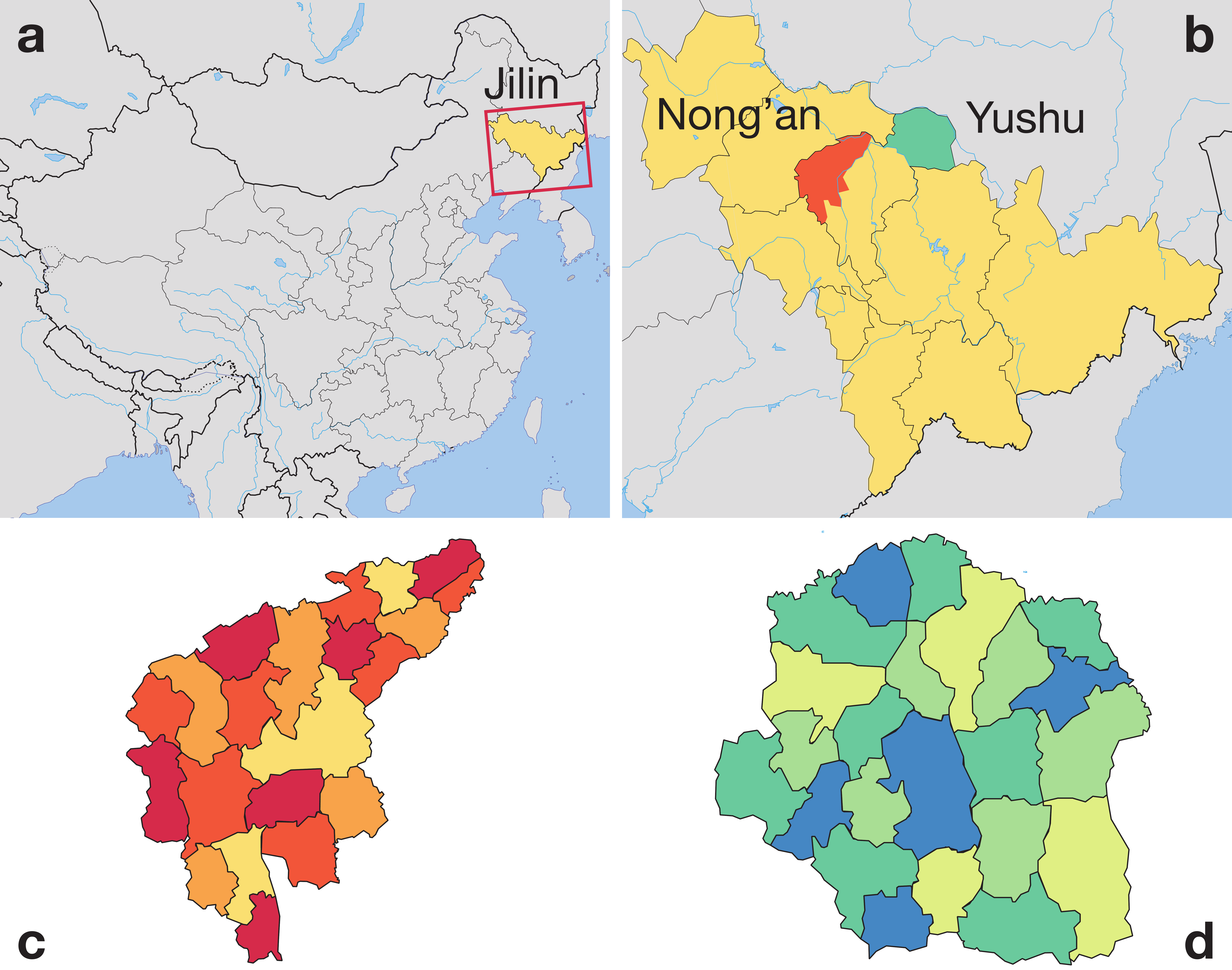} \caption{\textbf{Metapopulation susceptible-infected-recovered model for directly transmitted disease in Yushu City and Nong'an County.} Panel A shows the location of the Jilin province in northeastern China. Panel B is a blow-up of A showing the locations of our two data sets Yushu City and Nong'an County. Panel C (D) shows the 27 (22) divisions of Yushu City (Nong'an County). These divisions are administrative by construction, but also define the metapopulation we use in our modeling.
}
\label{fig:geo} 
\end{figure}

\begin{figure}[!ht]
\centering \includegraphics[scale=0.23]{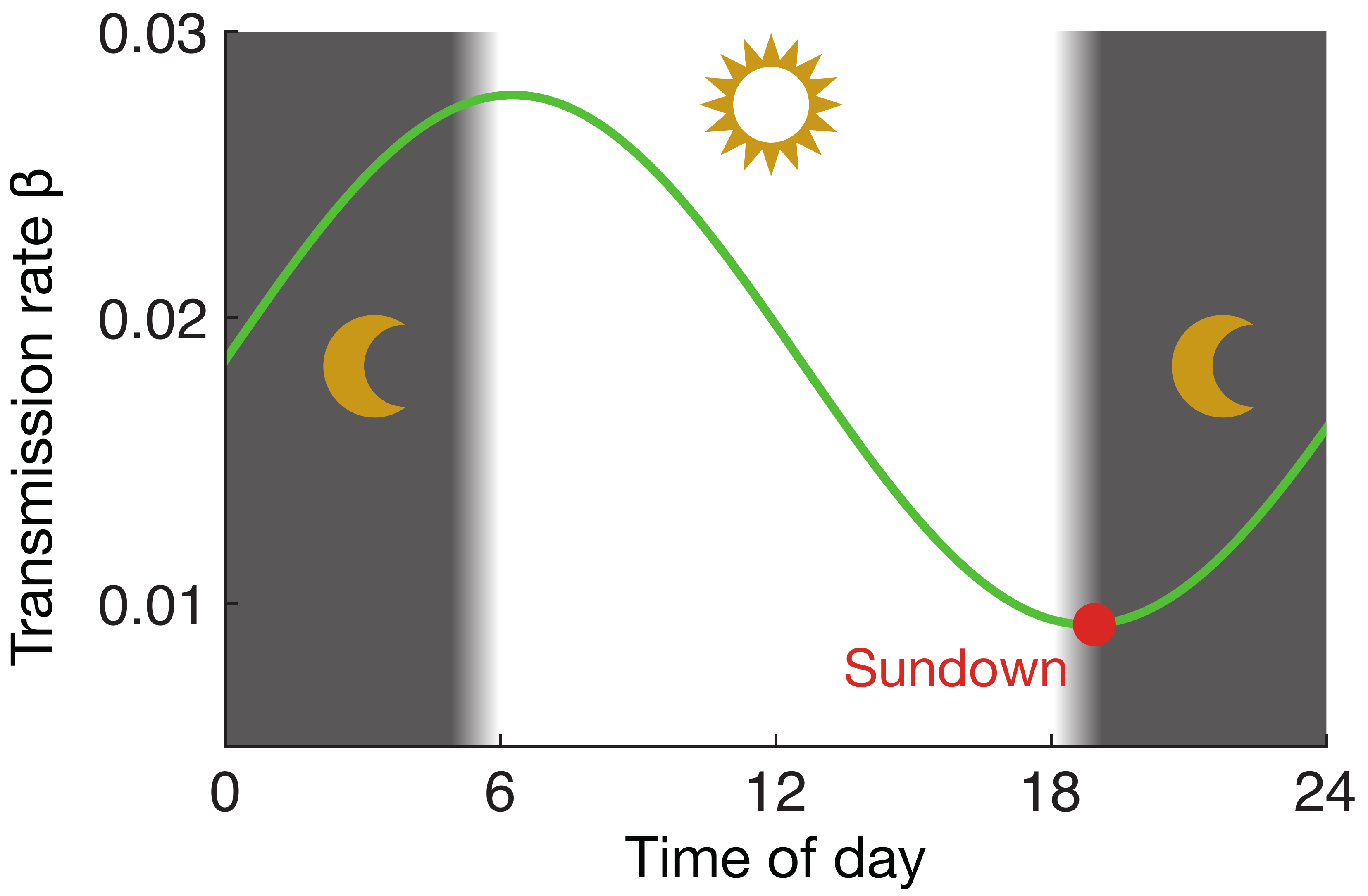} \caption{\textbf{Schematic representation of the transmission rate with respect to the circadian rhythms.}
To evaluate this temporal transmission rate, we divide a day into two stages, from sunrise to sundown and sundown to sunrise.
$\beta_t$ is the transmission rate in time $t$ with respect to the circadian rhythms [the average transmission rate $\beta_0=1/(2.25\times24)$, the controlling factor $C=0.5$, and $t_\text{max}=19/24$].
}
\label{figConcept} 
\end{figure}

As a basis for our study, we use real data of daily travel patterns as our substrate for interpersonal contacts. The data comes from mobility surveys from two counties (roughly the second level administrative division of China, after provinces)---Yushu City and Nong'an County---in China's northeastern Jilin province (Fig.~\ref{fig:geo}). From the data, we can see how many people that travelled between subdivisions of the counties at different times. This type of data lends itself well to a meta population type modeling, where one simulates assumes each location to be well-mixed and use the observed population flow to model the disease spreading between the locations~\cite{cooley2011role,dalziel2013human}. We use a susceptible-exposed-infectious-removed (SEIR) model with realistic parameters for influenza~\cite{fraser2009pandemic} and a susceptible-infectious-removed (SIR) model with realistic parameters for varicella~\cite{yang2016characterizing}. To incorporate the circadian patterns of the immune system, we use a sinusodial correction factor to the transmission rate parameters to model the circadian rhythms of the immune system. With this setup we investigate different scenarios like: the effect of including the circadian effects of the immune system (compared to modeling it as constant through the day) and the effect of shifting the mobility patterns relative to the immune system.

\section*{Methods}

\subsection*{Contact network from mobility data} 

We use mobility data from a public survey by two Chinese counties (Yushu City and Nong'an County)~\cite{du2019inter}.
These data sets give information about the spatiotemporal co-location of individuals and their movement between locations in the survey. Specifically, our data consist of one-week location records of anonymized cellular phone users across two counties in the Changchun area of China's northeastern Jilin province. These data sets were sampled between August 7, 2017 and August 13, 2017. The two counties consist of a mix of country side and urban areas are subdivided into 49 geographical divisions (Yushu City, with 1.18 million inhabitants, have 27 subdivisions, and Nong'an County, with 0.97 million inhabitants, in 22 divisions), corresponding to an administrative division~\cite{du2019inter,CCYearBook2017}.

We divide each county into metapopulations (henceforth \textit{locations}) corresponding to administrative divisions within a city as reported in the data (Fig.~\ref{fig:geo}). In this contact network, an edge denotes the flow of people between two locations and is weighted by the total number of users' movements in a specific hour of the data. In the raw data, the trips are reported as quadruplets $(a,i,j,t)$ meaning that individual $a$ travelled between locations $i$ to $j$ and arrived during the $t$'th hour of the data. From this data, we derive a series of 168 hourly mobility matrices $W_t$, one for each hour $t$ of the week. 
The $ij$-th entry of an hourly mobility matrix at hour $t$, $w_{ij,t}$, denotes the number of users traveling from location $i$ to location $j$ at time $t$ (at an hourly resolution).

For our analysis, we need a time-independent measure corresponding to \textit{strength}~\cite{Barrat3747} (the sum of a node's link weights) for our sequence of mobility matrices. To this end, we simply sum the weights of all the 168 matrices to get the strength
\begin{equation}
 s_i=\sum_{t=1}^{168} \sum_j w_{ij,t} .
\end{equation}

\subsection*{Epidemic model}

Basing disease simulations of large data sets of human mobility is challenging. Even though one has access to the trajectories of individuals, running individual based simulations is difficult~\cite{MEI201597}. How to compensate for missing data and how to scale up the simulations to actual city sizes makes this a questionable approach. Instead researchers rather use metapopulation models where the individuals are assumed to be well mixed within a metapopulation~\cite{balcan2009multiscale} at one segment of time. With metapopulation models, one still takes the flow of people and structures in geography and time into account~\cite{hanski1999metapopulation}. Many cases, including ours, one can argue that metapopulation models is modeling at the same level of detail as the original data set, since the sampled trajectories does not cover everyone moving in the area and this incompleteness hard to assess and compensate by adding simulated people.

Just like individual-based epidemic models, metapopulation models divide the population into classes (compartments) with respect to the disease and models the dynamics of the individuals within these compartments. We base or simulation on the work of Wesolowski \textit{et al.}~\cite{wesolowski2017multinational}, and thus considers four disease compartments: susceptible (individuals don not have, but could get, the disease), exposed (individuals who have got the disease, but cannot yet infect others), infective (individuals who have the disease and can spread it to susceptibles) and recovered (individuals who are immune or deceased and cannot get or spread the disease). For each location $i$, we denote the number of individuals in the three compartments at time $t$ by $S_{t,i}$, $E_{t,i}$, $I_{t,i}$ and $R_{t,i}$, respectively. In our simulations, time is effectively a discrete variable as the state of the system is reported at an hourly basis.

As epidemic outbreaks emerge, their transmission dynamics can vary much depending on the geographical locations of the initial infection~\cite{dalziel2013human,eubank2004modelling}. This maybe the most important insight from the recent, data-driven research on epidemics in urban populations~\cite{cooley2011role,dalziel2013human,herrera2016disease}. We chose a metapopulation model, that can capture such effects. In an entirely susceptible population at location $i$ at time $t$, an outbreak happens with probability: 
\begin{equation}
h(t,j)=\frac{\beta_t S_{j,t}(1-\exp[-\sum_k m_{j,k}^t x_{k,t}y_{j,t}] ) }{1+\beta_t y_{j,t}}
\end{equation}
where $\beta_t$ represents the transmission rate at hour $t$; $x_{k,t}$ and $y_{j,t}$ denote the fraction of the infected and susceptible populations at hour $t$ at location $k$ and location $j$, respectively, given by $x_{k,t} = \frac{I_{k,t}}{N_k} $ and $y_{j,t} = \frac{S_{j,t}}{N_j} $ where $N_k$ and $N_j$ are the population sizes of location $k$ and location $j$ according to the 2017 Changchun statistical yearbook~\cite{CCYearBook2017}.
Then we simulate a stochastic process introducing infections into completely susceptible metapopulations, in which $I_{j,t+1}$ is a Bernoulli random variable with probability $h(t,j)$. 

After disease introduction, the epidemic model is deterministic:
\begin{subequations}
\begin{eqnarray}
S_{j,t+1} &=& S_{j,t} -\frac{\beta_t S_{j,t} I_{j,t}}{N_j}\\ 
I_{j,t+1} &=& I_{j,t} + \frac{\beta_t S_{j,t} I_{j,t}}{N_j} -\gamma I_{j,t} \\
R_{j,t+1} &=& R_{j,t}+\gamma I_{j,t} 
\end{eqnarray}
\end{subequations}
or with an additional exposed class:
\begin{subequations}
\begin{eqnarray}
S_{j,t+1} &=& S_{j,t} -\frac{\beta_t S_{j,t} I_{j,t}}{N_j}\\ 
E_{j,t+1} &=& E_{j,t} + \frac{\beta_t S_{j,t} I_{j,t}}{N_j} -\sigma E_{j,t} \\
I_{j,t+1} &=& I_{j,t} + \sigma E_{j,t} - \gamma I_{j,t}\\
R_{j,t+1} &=& R_{j,t}+\gamma I_{j,t} 
\end{eqnarray}
\end{subequations}
where $\sigma$ is the incubation rate at which individuals move from the exposed to the infectious classes. $\gamma$ denotes the recovery rate at which people move from the infectious to the recovery classes.
An important parameter for the discussion is the basic reproduction number, $R_0$. It is defined as $R_0=\beta_t/\gamma$ and can be interpreted as the expected number of secondary infections if an infectious individual joins into a susceptible metapopulation. This reflects that the population in one location is well mixed. Everyone in a location at a given time has an equal probability to meet any other. The epidemics were simulated over all introduction locations and timings. In our setting, no matter when and where the epidemic begins, the outbreak will always reach the entire system, but the dynamics will vary with the parameter values. 

To model the circadian rhythm of the immune system, we modify the transmission rate by a sinusodial function~\cite{edgar2016cell,scheiermann2018clocking}:
\begin{equation}
\beta_t = \beta_0 \Big [1-C \cos [2\pi(t+\Delta t-t_\text{max})] \Big],
\end{equation}
where $t$ is the time of the day, $t_\text{max}$ is the time of the immune system maximum (both rescaled to the unit interval), $\beta_0$ is the average transmission rate over the day, $C\in[0,1]$ is the a factor controlling the strength of the circadian effects ($C=0$ meaning no effect, $C=1$ meaning the at the maximum of the immune system the transmission rate is zero (so the person is perfectly immune to the disease). 
See the illustration in Fig.~\ref{figConcept}. We include the parameter $\Delta t$ to model the effect of hypothetically shifting the travel patterns (for example by changing office hours or the time zone). If $\Delta t = 1$ hour then the travel patterns would be delayed one hour compared to the observed patterns.

\subsection*{Simulation setup and analysis}

In our model, outbreaks are characterized by five parameters---the transmission rate $\beta_0$, the incubation rate $\sigma$, the recovery $\gamma$, the time of the immune system maximum $t_\text{max}$, and the strength of circadian rhythm $C$. The temporal resolution is one hour. 
We set epidemiological parameters ($\beta_0$ , $\sigma$ and $\gamma$) with respect to two well-studied infectious diseases:
First, the 2009 H1N1 influenza pandemic (which we model by an SEIR model)---$\beta_0=1/(2.25\times24)$, $\gamma=1/(3.38\times24)$, $\sigma=1/(2.26\times24)$, and $R_0=1.5$~\cite{fox2017seasonality}. Second, the 2010 Taiwan varicella by SIR model---$\beta_0=1.50/24$, $\gamma=1/(5\times24)$ and $R_0=7.5$~\cite{yang2016characterizing}. During the time of the survey (mid-August), the time of the sunset (setting the immune system maximum) was 7:00 PM at both the data sets.

How to parameterize the circadian effect on the immune system, is still rather unknown in the medical literature. There are some results about herpes and influenza virus infections, respectively, with respect to viral replication volumes in the 24-hour circadian clock~\cite{edgar2016cell} pointing at the range $0.5<C<1$. We use three levels of the strength of the circadian impact on the immune system, spanning the entire range: high $C=1$, medium $C=0.5$ and low (or, rather, no) circadian effect $C=0$.

There are in total $16,464$ possible combinations of parameter values (which can be seen by multiplying the two $R_0$s (1.5 and 7.5), two cities, three circadian effect strengths, the 28 introduction timings (at the 8-th hour, 10-th hour, 16-th hour, and 21-th hour of each day of a week~\cite{du2018periodicity}), and 49 introduction sites. We make 20 runs for each parameter set, i.e.\ $329,280$ simulation runs with an initial infection seed and analyze these simulated time-series of the disease prevalence in two ways~\cite{du2018periodicity}:
\begin{enumerate}
\item We estimate the \textit{outward transmission risk} $\Gamma_{10\%}$ as the time (in hours) following an introduction into a location until $10\%$ of the locations have cases. 
\item We assess the \textit{introduction risk} $\chi$ as time (in hours) following introduction in a location until the focal location receives its first infection. This time is averaged over all introduction locations and all introduction times. 
\end{enumerate}

In a zeroth-order approximation, the transmission risk should be proportional to the location-to-location mobility and the transmission rate at that time of day. Let 
\begin{equation}\label{eq:transmission_potential}
 \Theta_{\Delta t} = \sum_{t,i,j} m_{i,j}^{t+\Delta t} \beta_t ,
\end{equation}
where $\Delta t$ denotes the time delay of mobility patterns. Then, to describe the effect of shifting the mobility patterns we measure the relative transmission risk as
\begin{equation}\label{eq:relative_transmission_potential}
 \Lambda_{\Delta t} =\frac{\Theta_{\Delta t}}{\Theta_{0}} ,
\end{equation}
where $\Theta_{0}$ represents the prevalence without travel times shifted.

\begin{figure}[!bpt]
\includegraphics[width=0.9\textwidth]{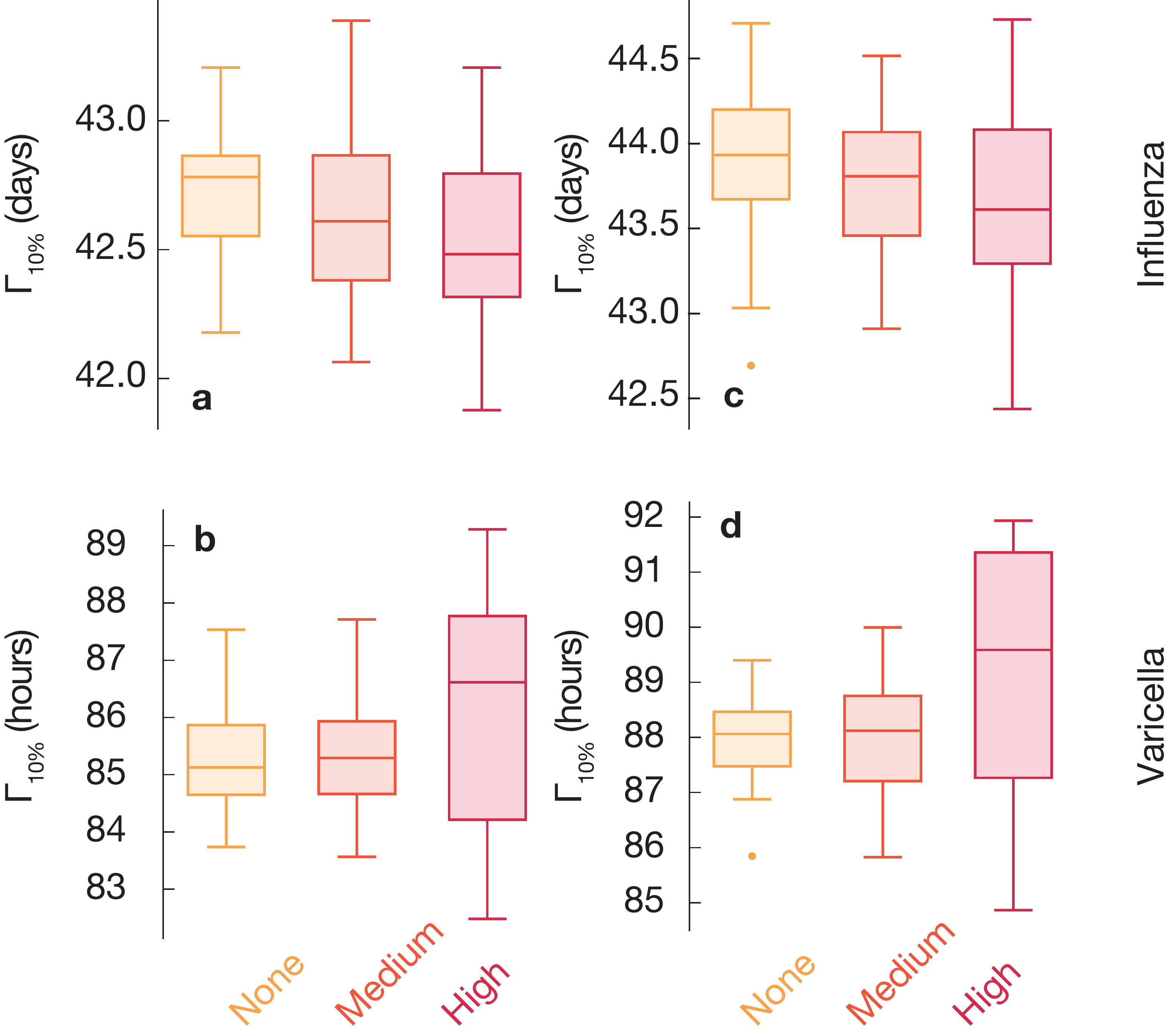} \caption{\textbf{Epidemic growth depends on circadian rhythms.} 
For each city, the time until 10\% of locations are infected ($\Gamma_{10\%}$), averaged over epidemics for each of the different introduction timings and sites. We compare simulations in settings with the time-dependent transmission rate at three different strengths of the circadian effects (high $C=1$, medium $C=0.5$ and no $C=0$ circadian effect).
Epidemic growth varies in different cities and levels of circadian rhythms.
Larger $\Gamma_{10\%}$ is observed with the time dependent transmission rate than without, decreasingly with higher $C$.
}
\label{figgrowthrates_Box} 
\end{figure}

\begin{figure}[!tbp]
\includegraphics[scale=0.3]{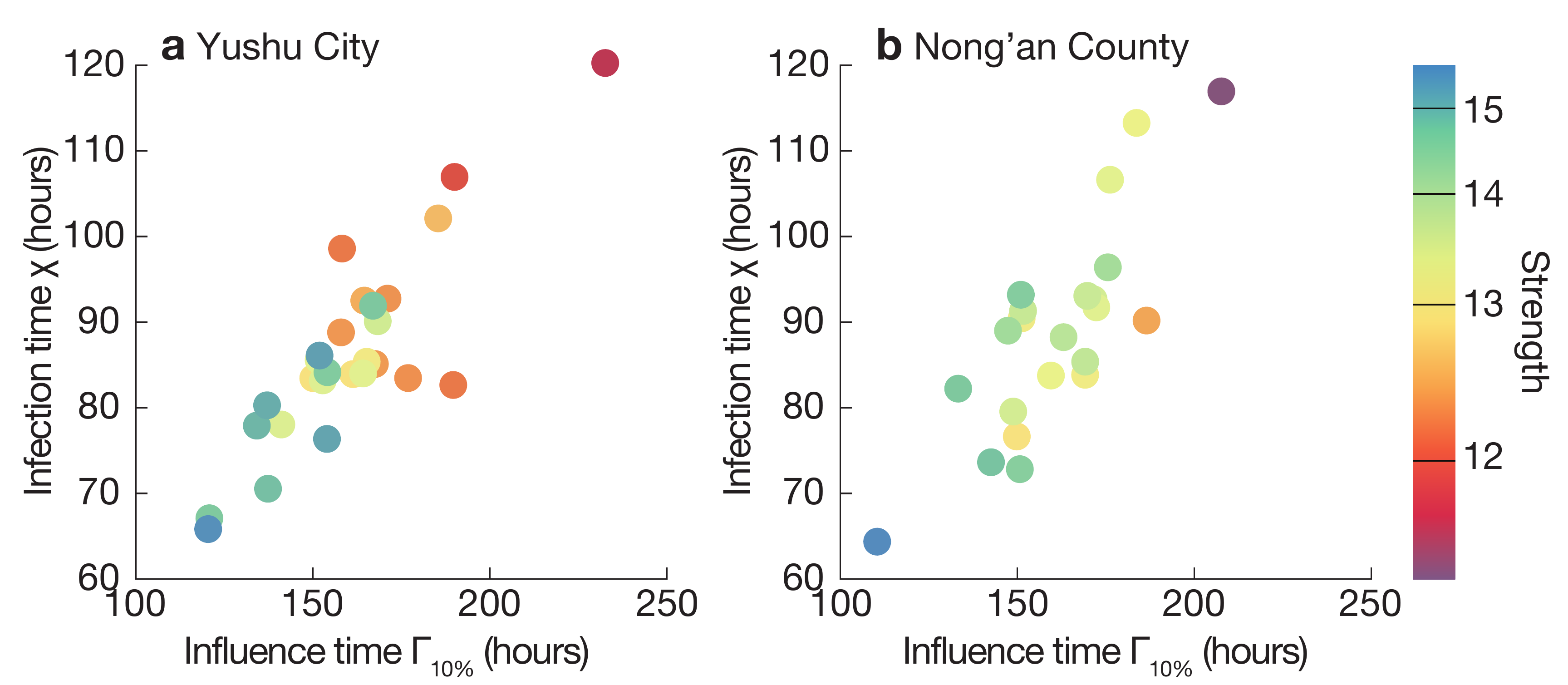} 
\caption{ 
\textbf{Risks of epidemic introduction and outward transmission correlate and increase with strength.} 
For each urban location in one of our two data sets, we estimate the risk of an outward transmission (x-axes) following an introduction into that location until $10\%$ of locations experience outbreaks. We average this value over different runs of the epidemic simulations for each of the $28$ different introduction times. The risk of introduction (y-axes) is estimated by the time since the introduction in another location until the focal location receives its first infection, averaged over all introduction location-time combinations. 
Colors correspond the strength (total mobility flow in and out from a location over seven days).
All these simulations assume an initial outbreak of no circadian effect and $R_0=1.5$. For the Yushu City data of panel (a), the inward and outward transmission risks, $\chi$ and $\Gamma_{10\%}$, are correlated with a Pearson's correlation coefficient of 0.83, and the risks correlate to strength with coefficients of $-0.85$ and $-0.75$, respectively.
}
\label{figTs10}
\end{figure}

\section*{Results}

\subsection*{Effects of including circadian rhythms in epidemic models}

Our first investigation concerns the epidemiological importance of introduction timing with and without explicit modeling of the circadian rhythm of the immune system. We measure the time following an introduction of the disease at a location until $10\%$ of the locations have infected people, i.e.\ the outward transmission risk $\Gamma_{10\%}$. As shown in Fig.~\ref{figgrowthrates_Box}, the epidemic spreads slowest in simulations without the circadian effects, fastest in the case of highest circadian dependence of the immune system. The effect is significant but rather weak. The observed pattern holds for both high and low $R_0$ but is somewhat more pronounced at low $R_0$. Furthermore, for each introduction location, we measure the introduction risk $\chi$---the number of hours following introduction in another location until the focal location receives its first infection over diverse introduction timings (see Fig.~\ref{figTs10}). The risk of introduction location to spread pathogens, as captured by $\Gamma_{10\%}$, seems to be anti-correlated with the strength of the locations (the total amount of in- and outflow of people), confirming observations in Ref.~\cite{du2018periodicity}.

In Fig.~\ref{figTs10}, we measure the network structural effect directly. We plot the correlations of three measures: strength (mobility flow into a location), $\Gamma_{10\%}$, and $\chi$. 
The three measures are highly correlated with each other. For example of simulations with no circadian effect and $R_0=1.5$ in the Yushu City, $\chi$ and $\Gamma_{10\%}$ are correlated with a Pearson's correlation coefficient of $0.83$, and the risks correlate to strength with coefficients of $-0.85$ and $-0.75$, respectively (all have $p<0.05$). Although not related to or much affected by the circadian effects, the centrality of locations, quantified by strength, is thus an important determinant of their role in the epidemics.

\begin{figure}[!h]
\includegraphics[width=0.95\textwidth]{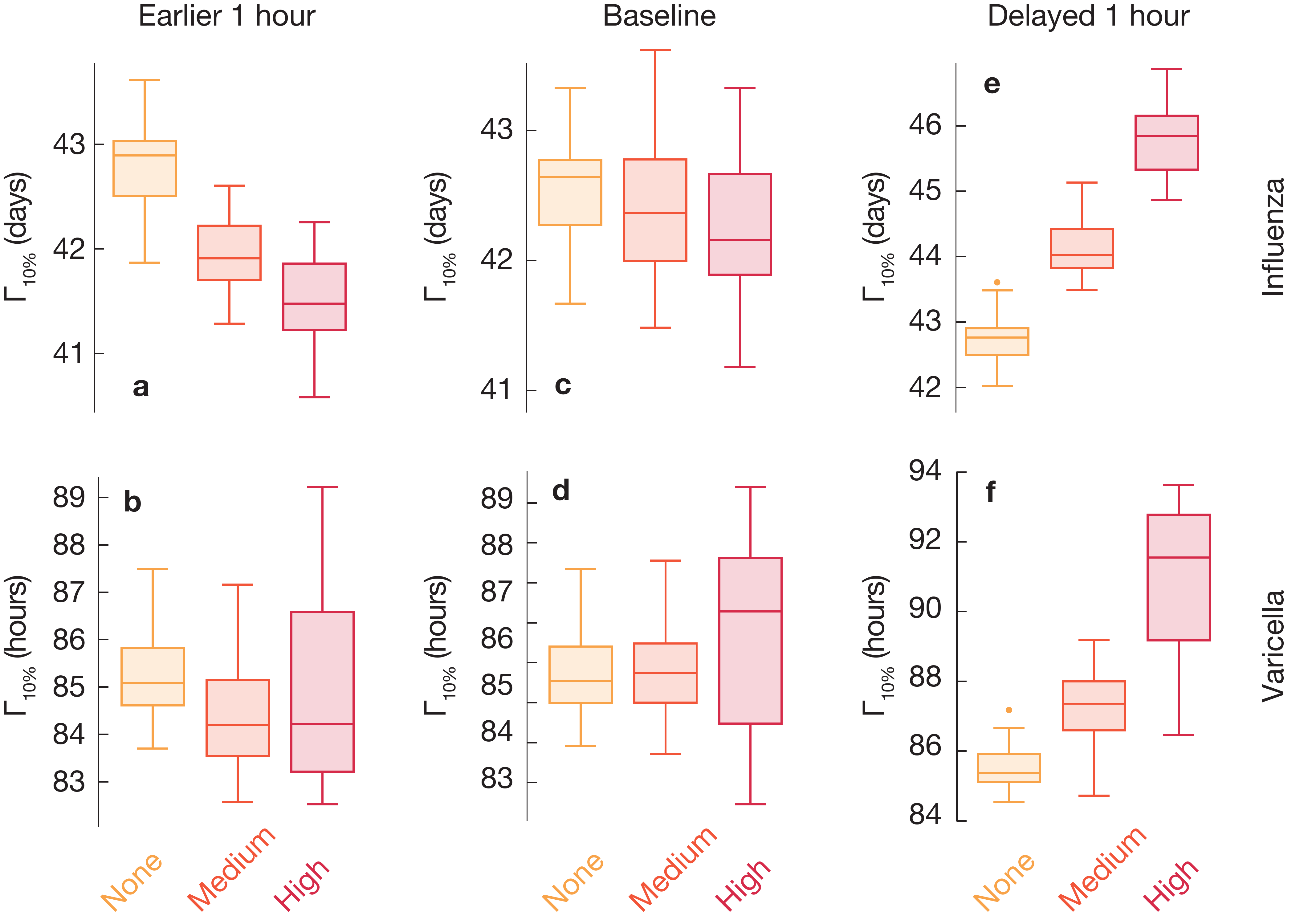}
\caption{\textbf{Epidemic growth if advancing or delaying mobility patterns one hour for the Yushu City data.}
Here we study epidemics by shifting mobility with one hour earlier (panels a and b) or later (e and f), in contrast with that without shifting (c and d). 
For each city, the time until 10\% of locations are infected ($\Gamma_{10\%}$) is averaged over epidemics for each of the different introduction times and sites.
We compare simulations in settings with the time dependent transmission rate characterized by three levels of circadian effects (high $C=1$, medium $C=0.5$ and $C=0$ with no circadian effect).
Epidemic growth varies in different levels of circadian rhythms.
Larger $\Gamma_{10\%}$ is observed with two patterns over $R_0$s: First, the time dependent transmission rate than without, increasingly with higher $C$, when mobility is earlier one hour or not. Second, the time dependent transmission rate than without, increasingly with higher $C$, when mobility is delayed one hour. These two patterns can be distinguish by the relative transmission risk (Table~\ref{tabMeasure}).
}
\label{figWorkYushu} 
\end{figure}

\begin{figure}[!ht]
\centering \includegraphics[width=0.95\textwidth]{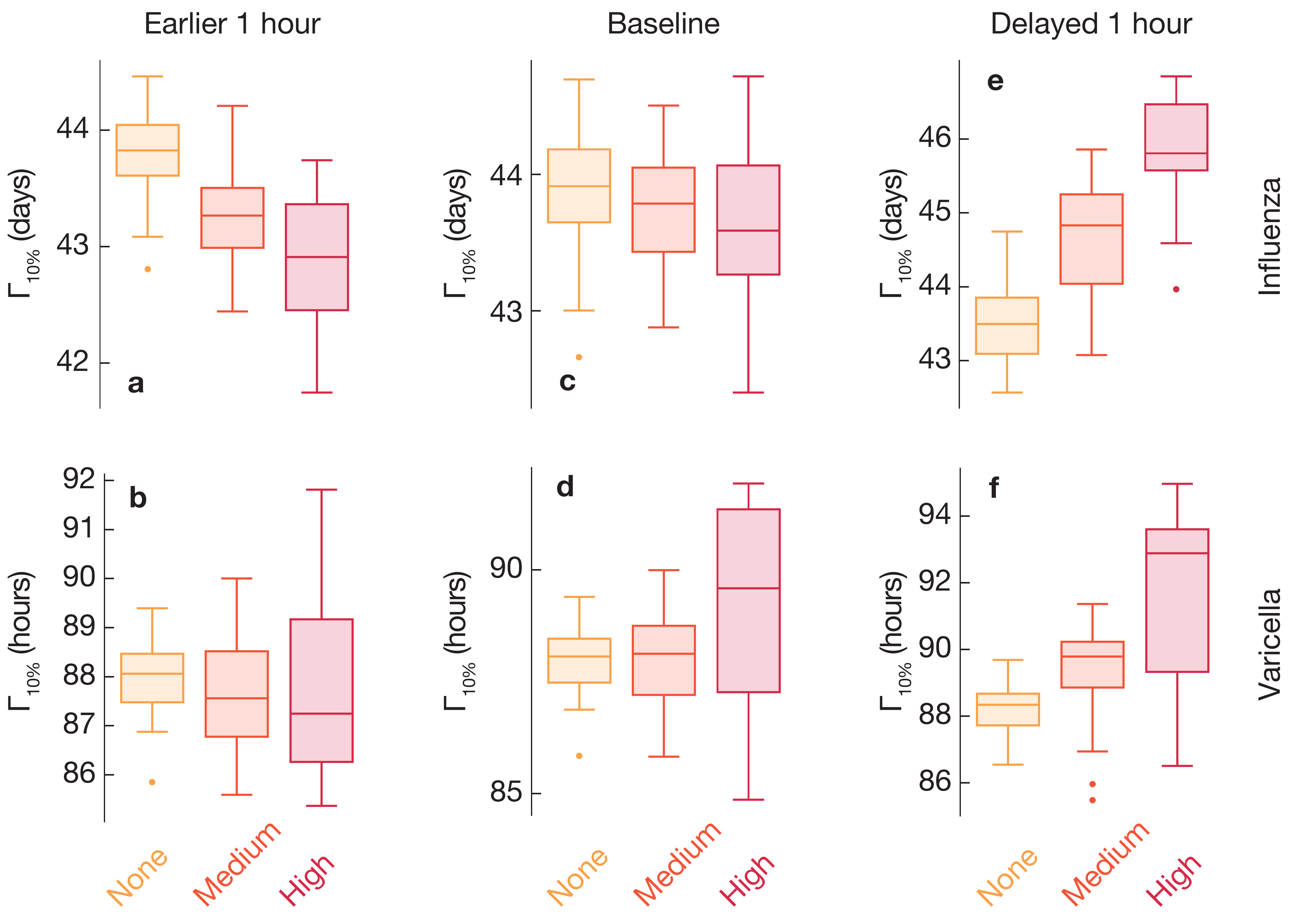} \caption{\textbf{Epidemic growth if advancing or delaying one hour for mobility in Nong'an County.} This figure corresponds to Fig.~\ref{figWorkYushu} but is for the data set from Nong'an County.
}
\label{figWorkNongan} 
\end{figure}

\begin{table}
\centering
 \caption{\textbf{Relative transmission risk when shifting the travel patterns.} We compare simulations in settings with the time dependent transmission rate characterized by three levels of circadian effects (high $C=1$, medium $C=0.5$ and $C=0$ with no circadian effect) in three scenarios by advancing, delaying mobility one hour or not. Taking mobility without shifting as baseline, we estimate the relative measure of ($\Lambda_{\Delta t}$) for each scenario to illustrate the potential transmission risk, from which we can distinguish trends with respect to circadian effects (see Figs.~\ref{figWorkYushu} and \ref{figWorkNongan}).}
\begin{tabular}{@{}llll@{}}
\toprule
 & & Earlier ($\Lambda_{-1}$) & Delayed ($\Lambda_{1}$) \\ \midrule
\multirow{3}{*}{Yushu City} & No & 0.934 & 1.066 \\
 & Middle & 0.971 & 1.030 \\
 & High & 1.057 & 0.943 \\ \midrule
\multirow{3}{*}{Nong'an County} & No & 0.937 & 1.063 \\
 & Middle & 0.975 & 1.025 \\
 & High & 1.062 & 0.938 \\ \midrule
\end{tabular}
\label{tabMeasure} 
\end{table}

\subsection*{Effects of shifting time of the travel patterns}

In the previous section, we established that there is a non-negligible effect of the circadian rhythms on the results of epidemic modeling. In this section, we look at the effects of the synchronicity between the immune system and the mobility patterns of people. Phrasing this in terms of parameter values, while in the previous section we were varying $C$ and kept $\Delta t$ constant, in this section we vary $\Delta t$ and keep $C$ constant.
In Figs.~\ref{figWorkYushu} and \ref{figWorkNongan}, we use a box-and-whiskers plots to show the effects of shifting the travel patterns forward or backwards in time. This effect is much bigger than the effect of tuning $C$. In particular, for the case of strong circadian rhythms $\Gamma_{10\%}$ gets delayed up to $10\%$, or seven hours, with one-hour delays of the travel rhythm of people $\Delta t = 1$ hour. (Relative changes are summarized in Fig.~\ref{tabMeasure}.) There is an asymmetry in this effect in that setting $\Delta t = -1$ gives the reverse effect, but of a lesser magnitude. This is a manifestation of what we mentioned above---that the effect of the coupling between travel patterns and the circadian rhythms of people is complex and trying to capture it by a general fudge factor might not be enough for an accurate modeling.

Furthermore, we observe that the effects of different diseases are different depending on whether $\Delta t$ is positive or negative. The SEIR simulation (with a lower $R_0$) of the influenza simulation makes the negative effects of advancing the travel patterns (negative $\Delta t$) much larger. For positive $\Delta t$ the effect is more similar.

\section*{Discussion and Conclusion}

We have investigated the effects of the coupling of the circadian rhythms of the immune systems and the daily travel patterns of city dwellers. In line with recent studies of urban epidemiology~\cite{hoen2015epidemic,ball2008network}, we used a metapopulation model based on real survey data that we parameterized to reflect two recent urban disease outbreaks of influenza and varicella~\cite{du2018periodicity,wesolowski2017multinational}. We modeled the circadian response of the immune system by multiplying the transmission rate by a sinusodial dependence of the time of day.

With this set-up, we found a small effect ($3\%$ difference or less) in including the circadian dependence of the immune systems. On one hand, possibly this effect is smaller than other sources of error, and could thus be neglected. On the other hand, it is possible that for other situations---other diseases, in other populations, at other locations and times of the year---this effect would be larger and less negligible. If possible, we thus recommend modelers to include the effect of the immune system's circadian dependence.

We also studied effects of shifting the clock time one hour relative to the sun. These effects could be larger (up to $10\%$) in the scenarios we simulated. Once again, this is specific to the travel data we studied (two urban regions in northeastern China, in August 2017). Probably, in even larger cities, or cities with yet more pronounced rush hours, these effects could be much larger. By shifting the time zone of a location, one could thus, in theory, slow down the disease propagation considerably, which would give more time for public health authorities to raise awareness of the disease and employ countermeasures. That much said, shifting the times of our daily lives would of course be a political process of great inertia. There is also the caveat that some of our routines are guided by the sun rather than the clock, so shifting the time zone is not the same as shifting the travel patterns. Clearly there is room for more research in the interface of chronobiology, epidemiology and data science (cf.\ Ref.~\cite{aledavood2018social}).

This work connects to a larger context of urban planning and social engineering. Earlier studies have found certain locations in cities to be extra important for epidemics~\cite{du2018periodicity}. This adds to the urban catalyst theory which explains the dynamics of cities as driven by certain points of interests~\cite{jacobs1961death,davis2009urban}---urban locations that undergoes a rapid development and redirects the flows of people. Such localities do not only have a positive effect on socioeconomic factors~\cite{geng2009effect,jiwei2006railway,papa2008rail}, but since they per definition have high strength (many people passing through them), they will also be key locations for the transmission of infections. By taking this into account in city planning, public facilities built to lower disease transmission would be efficient and perhaps cost effective~\cite{fernstrom2013aerobiology,chartier2009natural}.

\begin{backmatter}

\section*{Funding}
ZD would like to acknowledge funding from the Models of Infectious Disease Agent Study (MIDAS) program grant number U01 GM087719, Natural Science Foundation of Jilin Provincial Science and Technology Department (Grant No.\ 20180101332JC) and National Society Science Foundation of China (Grant No.\ 16BGL180).
PH was supported by JSPS KAKENHI Grant Number JP 18H01655.
The funders had no role in study design, data collection and analysis, decision to publish, or preparation of the manuscript.

\section*{Competing interests}
 The authors declare that they have no competing interests.

\section*{Author's contributions}
PH and ZD conceived the project. ZD carried out the simulations and analyses. PH and ZD wrote the paper.



\end{backmatter}
\end{document}